\documentclass[twocolumn,prl,aps,draft,nofootinbib]{revtex4}
\usepackage{psfig}

\def\lsim{\mathrel{\raise.3ex\hbox{$<$\kern-.75em\lower1ex\hbox{$\sim$}}}}
\def\gsim{\mathrel{\raise.3ex\hbox{$>$\kern-.75em\lower1ex\hbox{$\sim$}}}}

\begin{document}

%\hfill\vbox{
%\hbox{BUHEP-03-05}
%\hbox{MADPH-03-1321}
%\hbox{hep-ph/0405019}
%\hbox{}}

\title{Neutrino physics from new SNO and KamLAND data and future prospects}
\author{A.~B.~Balantekin$^1$, V.~Barger$^1$, D.~Marfatia$^{2,3}$,
S.~Pakvasa$^3$ and H.~Y\"{u}ksel$^1$ }
\vskip 0.3in
\affiliation{$^1$Department of Physics, University of Wisconsin, Madison, WI 53706, USA}
\vskip 0.1in
\affiliation{$^2$Department of Physics, Boston University, Boston, MA 02215,
  USA} 
\vskip 0.1in
\affiliation{$^3$Department of Physics and Astronomy, University of Kansas,
  Lawrence, KS, 66045  USA} 
\vskip 0.1in
\affiliation{$^4$Department of Physics and Astronomy, University of Hawaii, 
Honolulu, HI 96822, USA}
\begin{abstract}
We analyze the cumulative data from the latest SNO, KamLAND and 
other solar neutrino experiments in
the standard scenario of three oscillating active neutrinos. 
We determine the solar neutrino oscillation parameters  
and obtain new bounds on
$\theta_x$. We also place constraints on
 the fraction of oscillating solar neutrinos
that transform to sterile neutrinos with the $^8$B flux normalization 
left free.
Concomitantly, we assess the sensitivity of future data from the 
SNO and KamLAND experiments to $\theta_x$ and to the sterile 
neutrino content of the solar flux.

\end{abstract}

\maketitle

The SNO~\cite{Ahmad:2002jz} and KamLAND~\cite{Eguchi:2002dm} 
experiments have been crucial in selecting the Large Mixing Angle (LMA)
solution~\cite{Barger:2002at}, thereby solving the long-standing
solar neutrino problem. Additional KamLAND data~\cite{unknown:2004mb}
have narrowed the two-neutrino oscillation parameter space even 
further~\cite{unknown:2004mb,aliani}.
We perform a more general 
three-neutrino analysis of KamLAND and solar neutrino data including the
cumulative salt-phase SNO data announced recently~\cite{newsalt}. We 
refine the existing upper bound on $\theta_x${\footnote{We use the notation of 
Ref.~\cite{nurev} in which $\delta m^2_a$ and $\delta m^2_s$ are the 
atmospheric and solar mass-squared differences, and $\theta_a$, $\theta_s$ 
and $\theta_x$ are the mixing angles conventionally denoted by $\theta_{23}$,
$\theta_{12}$ and $\theta_{13}$, respectively.}}.
We also explore if future data from KamLAND and SNO can play an important role
in the study of neutrino physics beyond the determination of the
primary solar oscillation parameters.

One of the main goals of ongoing and planned neutrino experiments is 
a measurement of $\theta_x$, 
and if it is large enough, to determine if
$CP$ is violated in the neutrino sector~\cite{nurev}. 
Today, we know from the CHOOZ~\cite{Apollonio:1999ae} and
Palo Verde~\cite{Boehm:2001ik} experiments 
that $\sin^22\theta_x \leq 0.19$ at the 90\% C.~L. for
$\delta m^2_a=0.002$ eV$^2$; our analysis below yields 
$\sin^22\theta_x \leq 0.17$. Data from the
 K2K experiment have established an independent and consistent 
bound, $\sin^22\theta_x \leq 0.45$ 
for the same $\delta m^2_a$~\cite{Ahn:2004te}{\footnote{The aforementioned
 limits are quoted for two degrees of freedom.}}; 
further support that $\theta_x$ 
is small is obtained from Super-Kamiokande (SuperK) atmospheric 
data~\cite{nakaya}.
Long-baseline
experiments such as MINOS~\cite{minos} 
and the CERN to Gran Sasso (CNGS) 
experiments, ICARUS~\cite{icarus} and OPERA~\cite{opera}, 
will begin the hunt 
for $\nu_\mu \rightarrow \nu_e$ transitions resulting from a nonzero 
$\theta_x$ in the near future. 
Within five years of running they could have compelling 
evidence for such transformations or they
will strengthen the CHOOZ bound. In the meantime, however, there is a 
possibility that additional solar neutrino 
data may provide guidance on the size of 
$\theta_x$. A constraint from solar neutrino data is independent of 
$\delta m^2_a$ so long as it is much larger than $\delta m^2_s$.
This is especially important because the values of $\delta m^2_a$ from 
the SuperK collaboration's analyses have shifted with additional
data and refinements in the analyses (in
quite a narrow range which, however, sensitively affects conclusions about
 the size of $\theta_x$); compare the results from a zenith-angle 
analysis~\cite{hayato} and from an $L/E$ analysis~\cite{shiozawa}.
If $\delta m^2_a$ turns out to be smaller than $0.001$ eV$^2$, then the
CHOOZ bound will be inoperable, and solar data will provide the most
stringent bound on $\theta_x$; even MINOS and the CNGS experiments
will not do better. Although we have no reason to
believe that this will be the case, 
we mention this as a hypothetical 
possibility under which solar/KamLAND data provide the best bound
on $\theta_x$. After all, the K2K experiment confirms the $\delta m^2_a$
values from SuperK at the 2$\sigma$ C.~L.~\cite{k2k}.

More realistically, we investigate if future solar data 
can improve on the CHOOZ bound for the $\delta m^2_a$ values that are
consistent with SuperK and K2K.

Another unresolved issue is whether solar neutrinos oscillate
into sterile species~\cite{cirelli}. We know from solar data that the 
possibility
that solar neutrinos oscillate exclusively to sterile states is excluded 
at 7.6$\sigma$~\cite{nurev}. However, it is easily conceivable that
solar $\nu_e$ oscillate into both active and sterile neutrinos. The latter
scenario is not satisfactorily constrained at present, and significant
improvement in this direction is unlikely in the near 
future~\cite{Barger:2001zs}. We evaluate how future SNO and KamLAND data
may confirm and somewhat improve existing bounds on a sterile fraction in the
solar flux with minimal dependence on the Standard Solar Model (SSM) 
and without resort to
involved global analyses of strongly correlated datasets from many experiments.

All the $^3$He proportional counter tubes or neutral current detectors
 are installed and are taking data for the
third phase of the SNO experiment. The future 
NC measurement is expected to have
an overall uncertainty (statistical and systematic uncertainties 
combined) 
of about 6.4\%. At the same time an improved CC integrated flux 
measurement 
will be made with an expected overall uncertainty of about 5.5\%. 
To a good approximation, these measurements will be uncorrelated with
previous measurements and with each other. We use these expectations 
in our analyses.

In the analysis of the latest KamLAND data we take into account the fact that 
some of the reactors were nonoperational by using the expected number of 
nonoscillated events given in Fig.~1 of Ref~\cite{unknown:2004mb}.

%For future KamLAND data, we suppose that KamLAND selects $\delta m^2_s$
%in the Large Mixing Angle (LMA) region below $10^{-4}$ eV$^2$~\cite{Barger:2002at} with 2.5 times more events 
%than the present cumulative total in Ref.~\cite{unknown:2004mb} 
%assuming that all reactors are operational. 
%Throughout, we simulate KamLAND spectrum data for the best-fit parameters from
%our current analysis,
%$\delta m^2_s=8.3\times 10^{-5}$ eV$^2$ and 
%$\tan^2 \theta_s=0.42$.

We employ the SSM~\cite{Bahcall:2004fg} in our analyses, 
but treat the $^8$B flux normalization as a free parameter throughout.

\vskip 0.1in
\noindent
{\bf Sensitivity to $\theta_x$}:

\begin{figure}[t]
  \begin{center}
\psfig{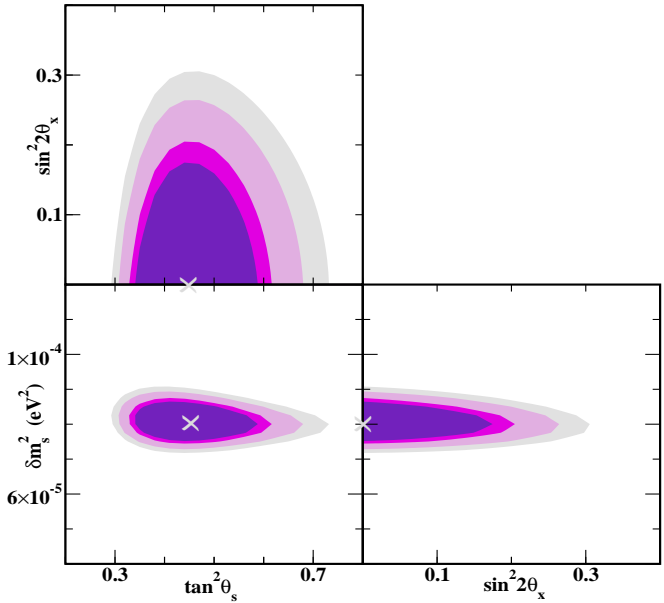}
%\bigskip
\caption[]{The 90\% C.~L., 2$\sigma$, 99\% C.~L. and 3$\sigma$ allowed 
regions from a combined three-neutrino 
fit to CHOOZ, KamLAND and solar neutrino data. The
best-fit point $\delta m^2_s=8\times 10^{-5}$ eV$^2$, 
$\tan^2 \theta_s=0.45$ and $\sin^2 2\theta_x=0$ is marked with an ``X''. 
In the analysis, the $^8$B flux was a free parameter.}
    \label{fig:allow}
  \end{center}
\end{figure}

\begin{figure}[h]
  \begin{center}
\psfig{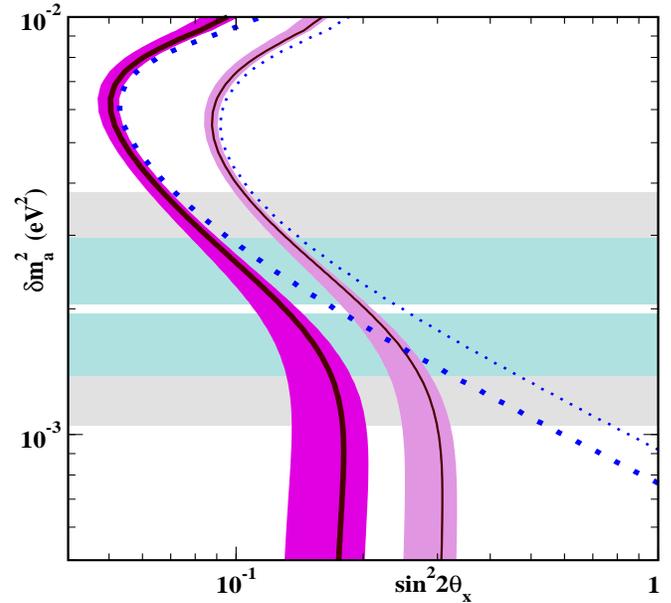}
%\bigskip
\caption[]{Estimates of how future SNO data will affect bounds
on $\theta_x$. The shaded curved bands depict
 the effect of future SNO 6.4\% NC and 
5.5\% CC measurements (whose central values 
lie within their current 1$\sigma$ values) 
on bounds from all existing CHOOZ, KamLAND and solar
neutrino data. The  thick (thin) solid curves
are the 90\% C.~L. (3$\sigma$)  
bounds from current CHOOZ, KamLAND and solar data.
The dotted curves are the 
corresponding CHOOZ bounds. 
The horizontal shaded regions encompass the values of 
$\delta m^2_a$ favored by 
SuperK atmospheric data at the 90\% and 99\% C.~L.~\cite{hayato}. 
The $^8$B flux normalization is a free parameter
in our analyses. 
}
    \label{fig:t13}
  \end{center}
\end{figure} 

For the $\nu_e$ survival probability in the three-neutrino framework, we use
the standard modification of the two-neutrino survival probability as derived
in Ref.~\cite{Fogli:2000bk}.

The regions of parameter space allowed by existing CHOOZ, KamLAND
and solar data are shown in Fig.~\ref{fig:allow}. 
%We find excellent agreement
%with the results of Ref.~\cite{malt}. 
%Note that while the best-fit value
%of $\sin^2 2\theta_x$ is 0.03, the 
%significance of this nonzero value is only 0.05$\sigma$.

The effect of how future data from the SNO experiment will
impact our knowledge of $\theta_{x}$ is comprehensively represented in
Fig.~\ref{fig:t13}. The figure clearly suggests that
future SNO data will not have a significant
impact on existing bounds, especially for $\delta m^2_a$ values relevant to
atmospheric neutrino oscillations. 

%KamLAND data may lead to some improvement
%provided $\delta m^2_a$ is close to 0.001 eV$^2$; see the bottom panel.
%It appears that future measurements by SNO and KamLAND may
%provide only a marginal improvement in the limits for 
%$\theta_x$.

%

\vskip 0.1in
\noindent
{\bf Sensitivity to sterile neutrinos}:

\begin{figure}[t]
  \begin{center}
\psfig{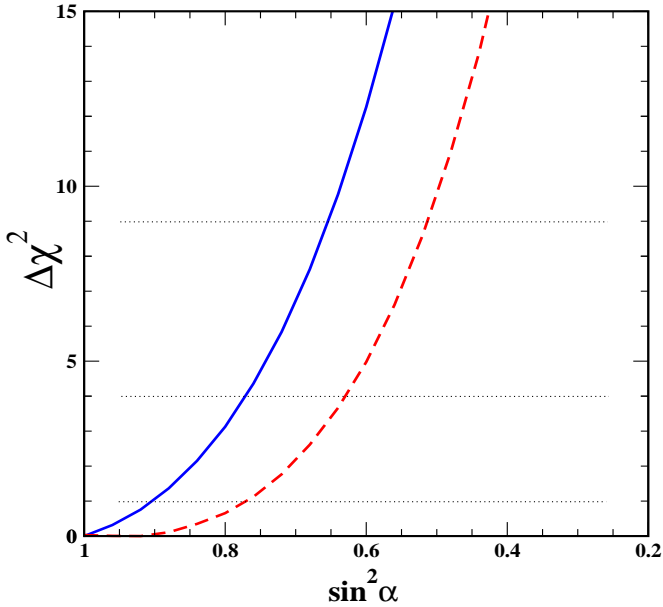}
%\bigskip
\caption[]{$\Delta \chi^2$ vs $\sin^2 \alpha$ from analyses of 
all solar and KamLAND data (solid),
and only SNO and KamLAND data (dashed), with the $^8$B flux free in both 
analyses. 
From bottom to top, 
the horizontal dotted lines indicate the $\Delta \chi^2$ values 
corresponding to $1\sigma$, $2\sigma$ and $3\sigma$.}
    \label{fig:ster}
  \end{center}
\end{figure}

In a scenario in which oscillations to sterile neutrinos are allowed, 
the fraction of oscillating neutrinos that transform to active neutrinos 
is (in terms of quantities measured by SNO)~\cite{Barger:2001zs},
\begin{equation}
\sin^2 \alpha= {\Phi_{NC}-\Phi_{CC} \over \Phi_{^8B} - \Phi_{CC}}\,.
\end{equation}
%where $\Phi_{^8B}=\beta \Phi_{SSM}$ for the SSM, with 
%$\Phi_{SSM}=5.79\times 10^6$ cm$^{-2}s^{-1}$ and $\beta=1\pm 0.23$. 
The current constraints on $\sin^2 \alpha$ are shown in Fig.~\ref{fig:ster}. 
%The most stringent bound
%from all available solar and KamLAND data with the $^8$B flux constrained
%by the SSM value is $\sin^2 \alpha \geq 0.94$ (0.7) at 
%1$\sigma$ (3$\sigma$). For our present purpose the SSM $^8$B flux prediction
% is not robust enough, and so these
%bounds are not conservative. 
The most
stringent bound from all available solar and KamLAND data is
$\sin^2 \alpha \geq 0.91$ (0.65) at 1$\sigma$ (3$\sigma$). 
Our estimates are conservative since the
$^8$B flux normalization is left free in the analyses. 

%\begin{figure}[h]
%  \begin{center}
%\psfig{file=ssm.eps,width=8.75cm,height=8cm}
%\bigskip
%\caption[]{Iso-$\sigma_{\sin^2 \alpha}/\sin^2 \alpha$ (solid) lines and 
%iso-$\sin^2 \alpha$ (dotted) lines for future 6.4\% $\Phi_{NC}$ and 
%5.5\% $\Phi_{CC}$ measurements from SNO. In the left (right) panel, the 
%SSM value (the SSM value with half the uncertainty) is adopted for  
%$\Phi_{^8B}$. The ``+'' signs mark the current central values of 
%$\Phi_{NC}$ and $\Phi_{CC}$ measured by SNO.}
%    \label{fig:ssm}
%  \end{center}
%\end{figure}

\begin{figure}[tbh]
  \begin{center}
\psfig{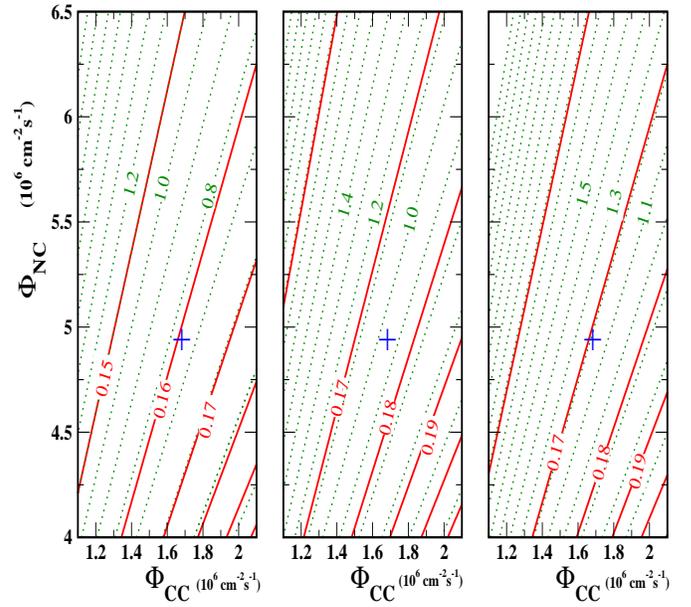}
%\bigskip
\caption[]{Iso-$\sigma_{\sin^2 \alpha}/\sin^2 \alpha$ (solid) lines and
iso-$\sin^2 \alpha$ (dotted) lines for future 6.4\% $\Phi_{NC}$ and
5.5\% $\Phi_{CC}$ measurements from SNO. $\Phi_{^8B}$ is
  obtained from Eq.~(\ref{eq:pee}) and $P_{ee}$ is determined by KamLAND with
7\% uncertainty. From left to right, the three panels are for three
possible measurements, $P_{ee}=0.28$,
0.33 and 0.38, respectively. The ``+'' signs mark the current central values
of
$\Phi_{NC}$ (=4.94) and $\Phi_{CC}$ (=1.68) measured by SNO.}
    \label{fig:pee}
  \end{center}
\end{figure}

Our knowledge of $\sin^2\alpha$ can be refined if we can
observationally
infer the $^8$B flux produced in the Sun. We
now describe such a method.

The KamLAND experiment which detects ${\bar\nu}_e$ from surrounding nuclear
reactors will determine the solar oscillation parameters
to 10\% precision independently of solar physics.
These parameters can be used
as inputs in analyses of SNO data to extract the average $\nu_e$
survival probability measured by SNO. The solar flux can be obtained via
\begin{equation}
\Phi_{^8B} = \Phi_{CC}/P_{ee}\,,
\label{eq:pee}
\end{equation}
where $P_{ee}$ is the average survival probability of $\nu_e$ at SNO.
It has been shown in Ref.~\cite{Bahcall:2002zh}
that with a few years of KamLAND data, $P_{ee}$
should be known to about 7\% for parameters in the LMA region obtained from
solar data.
Although
matter effects in the Sun depend on the active-sterile admixture,
for the oscillation parameters and sterile fraction allowed by current
data, they have little effect on $P_{ee}$.

The dotted lines in Fig.~\ref{fig:pee} are iso-$\sin^2 \alpha$ lines and the 
solid lines are iso-$\sigma_{\sin^2 \alpha}/\sin^2 \alpha$ lines, or 
lines with
the same fractional uncertainty in the $\nu_{\mu, \tau}$ content at
1$\sigma$. 
%The left (right) panel is for $\beta=1 \pm 0.23$ 
%($\beta=1 \pm 0.12$). 
Although $\sin^2 \alpha>1$ values are unphysical,
they are experimentally obtainable
since $\Phi_{NC}$ could be measured to be higher than $\Phi_{SSM}$. 
The figure should be interpreted as follows: Each point marks the central 
values of the $\Phi_{NC}$ and $\Phi_{CC}$ measurements with 6.4\%
and 5.5\% uncertainties, respectively. The solid line passing through
each point gives the corresponding 
$\sigma_{\sin^2\alpha}/\sin^2 \alpha$. 
Since the expected uncertainties on $\Phi_{NC}$ and $\Phi_{CC}$ are 
incorporated in the solid lines, one should not plot the measurements
with their uncertainties to read-off the envelope of 
$\sigma_{\sin^2\alpha}/\sin^2 \alpha$.
%Note that since the
%solid (dotted) lines are essentially vertical (horizontal),
%$\sigma_{\sin^2\alpha}/\sin^2 \alpha$ will be 
%determined by the $\Phi_{CC}$ measurement
%while $\sin^2 \alpha$ will be determined by the $\Phi_{NC}$ measurement.
  
%For future measurements of $\Phi_{NC}$ and $\Phi_{CC}$ with central values 
%that are consistent
% with present determinations, we expect $\sigma_{\sin^2 \alpha}/\sin^2 \alpha$ 
%to be about 32\%, which is almost identical to the existing
%uncertainty; see the dotted line in Fig.~\ref{fig:ster}. 
%Thus, we concur with the conclusions of Ref.~\cite{Barger:2001zs} 
%that because of the large
%uncertainty in the SSM $^8$B flux prediction, even a precise NC measurement
%by SNO will not permit a meaningful determination of the sterile neutrino 
%fraction.
%If the uncertainty in $\beta$ 
%is improved by a factor of 2 so that, $\beta= 1 \pm 0.12$, 
%$\sigma_{\sin^2 \alpha}/\sin^2 \alpha$ will be reduced to about 19\%; see the
%right panel of Fig.~\ref{fig:ssm}.   

%Another way to reduce $\sigma_{\sin^2 \alpha}/\sin^2\alpha$ is to 
%observationally
%infer the total $\nu_e$ flux produced in the Sun, such as by the method we now
%describe.

In  Fig.~\ref{fig:pee}, from left to right, 
we show our expectations for $\sigma_{\sin^2 \alpha}/\sin^2 \alpha$
for $P_{ee}=0.28$, 0.33 and 0.38, all with 7\% uncertainties. 
Since both the solid and dotted lines have slopes higher
than 2.5, both $\sigma_{\sin^2\alpha}/\sin^2 \alpha$ and $\sin^2 \alpha$ 
will have greater sensitivity to the value of $\Phi_{CC}$ than to the value of
$\Phi_{NC}$. We conclude
that $\sigma_{\sin^2 \alpha}/\sin^2 \alpha$ will be known to 16--17\%. These 
projections are comparable with existing bounds as represented by the
dashed line of Fig.~\ref{fig:ster}. 

Since these expectations are based only on future SNO and KamLAND data,
they are conservative. Further improvement can be achieved by combining
with other solar data. Joint analyses of solar data are dictated
by the paucity of the data. 
With the future availability of larger datasets it will be worthwhile 
to perform more definitive analyses of data from experiments which do not have 
correlations with each other (such as SNO and KamLAND).

\vskip 0.1in
\noindent
{\bf Conclusions}:

In a three-neutrino framework, our analysis of all existing KamLAND, CHOOZ 
and solar neutrino data yields 
\[
\delta m^2_s = 8.0^{+0.7}_{-0.6} \times 10^{-5} {\rm {eV}}^2\,,\ \ \ 
\tan^2 \theta_s=0.45^{+0.17}_{-0.12}\,,
\]
 where the uncertainties are at 
the 2$\sigma$ C.~L.
Current bounds on $\theta_x$ are significantly improved for lower values of $\delta m^2_a$ favored by SuperK. 
For the SuperK best-fit 
$\delta m^2_a=0.002$ eV$^2$, the 
CHOOZ upper limit is slightly improved by KamLAND and solar data to
\[
\sin^2 2\theta_x \leq 0.13\  (0.20)
\]
at the 90\% C.~L. (3$\sigma$). 

The fraction of solar neutrinos oscillating into active neutrinos is
greater than (0.91) 0.65 at 1$\sigma$ (3$\sigma$) from all existing
solar and KamLAND data.

%Future 
%SNO measurements will not be sensitive to $\theta_x$ and existing bounds
%will essentially remain unaffected. 
%While KamLAND data alone are not sensitive
%to $\theta_x$~\cite{Barger:2000hy} either, 
%by measuring the solar mass-squared difference precisely, 
%it may become possible to slightly
%strengthen the constraint if $\delta m^2_a$ turns
%out to be closer to 0.001 eV$^2$ than presently expected. 
A substantially improved constraint
on $\theta_x$ from future SNO data should not be anticipated
unless $\delta m^2_a$ is at the lower edge of what SuperK atmospheric data
prefer (in which case, the CHOOZ data are not very constraining).

With future SNO and KamLAND data alone, it will be
possible to know the fraction of solar neutrinos transforming to active 
species to a precision of 16--17\% at 1$\sigma$. 
This will be an important confirmation of existing bounds 
because the SNO and KamLAND datasets 
are completely uncorrelated with each other. A nonnegligible 
sterile neutrino component in the solar flux incident on the earth
will remain a possibility.

\vskip 0.1in
{\it Acknowledgments}:
We thank A.~McDonald for sparking our interest and for useful 
discussions and comments on the manuscript. 
This research was supported by the U.S.~DOE  
under Grants No.~DE-FG02-95ER40896, No.~DE-FG02-91ER40676 and
No.~DE-FG03-94ER40833, by the
U.S.~NSF under Grant No.~PHY-0244384,
and by the Wisconsin Alumni Research Foundation. 

\end{document}